\documentclass[12pt]{article}
\usepackage{amsmath}
\usepackage{graphicx,amsmath}
\usepackage{amssymb,amsthm} %
\usepackage{subfig}
\usepackage{array}
\usepackage{multirow}
\usepackage{tikz}
\usepackage[titletoc,title]{appendix}

\usepackage[modulo]{lineno}
\linenumbers
\usepackage[affil-it]{authblk}

\begin{document}

\title{Fast approximate delivery of fluence maps: the single map case}

\author{David Craft%
  \thanks{Electronic address: \texttt{dcraft@partners.org}; Corresponding author}}
\affil{Department of Radiation Oncology, Massachusetts General Hospital and Harvard Medical School, Boston, MA 02114, USA}

\author{Marleen Balvert}
\affil{Department of Econometrics and Operations Research/Center for Economic Research (CentER), Tilburg University, PO Box 90153, 5000 LE Tilburg, The Netherlands}

\date{Dated: \today}

\maketitle

\begin{abstract}
  In this first paper of a two-paper series, we present a method for optimizing the dynamic delivery of fluence maps in radiation therapy.
  For a given fluence map and a given delivery time, the optimization of the leaf trajectories of a multi-leaf collimator to approximately form
  the given fluence map is a non-convex optimization problem. Its general solution has not been addressed in the literature, despite
  the fact that dynamic delivery of fluence maps has long been a common approach to intensity modulated radiation therapy.
  We model the leaf trajectory and dose rate optimization as a non-convex continuous optimization problem and solve it by an
  interior point method from randomly initialized feasible starting solutions. We demonstrate the method on a fluence map from a prostate
  case and a larger fluence map from a head-and-neck case. 
  While useful for static beam IMRT delivery, our main motivation for this work is the extension to the case of
  sequential fluence map delivery, i.e. the case of VMAT, which is the topic of the second paper.
\end{abstract}

\section{Introduction}

The fast delivery of a fluence map (also sometimes referred to as an intensity map) has received some attention over 
the years \cite{crooks,engel,klink}
but the following remains an unsolved problem in general: for a given allotted delivery time, determine a set of leaf trajectories and dose rates versus time to best recreate a given fluence map, given machine characteristics such as dose rate restrictions 
and maximum leaf speed. Since the efficient delivery of fluence maps is at the heart of intensity modulated radiation therapy (IMRT) 
and volumetric modulated arc therapy (VMAT), we have returned to this basic question in order to improve IMRT and in particular, VMAT.
In the first of the two papers in this set we focus on the single fluence map case and in the second paper we use the single map
findings in order to address the delivery of sequential fluence maps, which is the case of VMAT.

Fluence map delivery by a multi-leaf collimator (MLC), the defining hardware component
of IMRT, is done in either a step-and-shoot or dynamic fashion. 
In step-and-shoot delivery, the MLC leaves are moved into position while the beam is off, then the beam is turned on for delivery, 
and then this process is repeated for each segment shape to deliver \cite{xia}. 
In dynamic delivery, the beam is on while the leaves are moving, painting out the fluence map \cite{yu1995}.

We focus on the dynamic delivery of a fluence map since our ultimate goal is 
VMAT, which inherently utilizes dynamic delivery. We are motivated by the simple question: {\em given a fluence map and a 
fixed delivery time, what is the best we can do in terms of matching that fluence map?} For the case
of infinite leaf speed and assuming the leaves can move across the field with no tip gap (i.e. fully closed), the sliding window algorithm with maximum dose rate is known to be optimal \cite{bortfeld94,stein94}. 
However, for finite leaf speed the optimal delivery will in general not be a trajectory where all the leaf pairs slide across 
the entire field. For example, for a uniform field, the optimal delivery will be to set the leaves open at the field boundaries
and irradiate at maximum dose rate. Likewise, for row-wise unimodal fields, a close-in (or open-out) technique
will be faster than a sweeping window technique.

These simple ideas, while useful for thinking about optimal dynamic delivery, are not enough to solve the fluence map delivery
problem in general. The difficulty is that the rows of a fluence map, while independent from each other regarding leaf
motions (assuming that interdigitation is allowed and that we do not try to reduce tongue-and-groove effect), are coupled
via the time-varying dose rate, which applies simultaneously to the entire field. 
An optimal dose rate versus time for one leaf row considered in isolation
will in general be unique to that row. We therefore model the problem as a constrained optimization problem and solve for all 
leaf motions and the dynamic dose rate simultaneously.

\section{Materials and Methods}
To isolate the problem to fluence map delivery, we assume the standard two-step approach to IMRT planning. That is, we assume that
the IMRT optimization problem is solved by first optimizing for the fluence maps and then applying a leaf sequencing algorithm for the 
delivery of those fluence maps \cite{jelen}. For this work, we ignore the first optimization and simply  
assume that the fluence maps are given as a result of an IMRT optimization. 
We also assume that the treatment delivery device, a linear 
accelerator (linac) equipped with an MLC, has the following characteristics:
\begin{itemize} 
\item Continuously variable dose rate up to some maximum level.
\item Continuously variable MLC leaf speeds up to some maximum level.
\end{itemize}

We do not model the jaws of the linac and we ignore leaf transmission. We also assume leaves can shut fully and move in this fully shut position.  In the Discussion section we describe how to include these features.

\begin{figure}[!htbp]
\centering
\includegraphics[trim=0 300 0 50,clip,width=12cm]{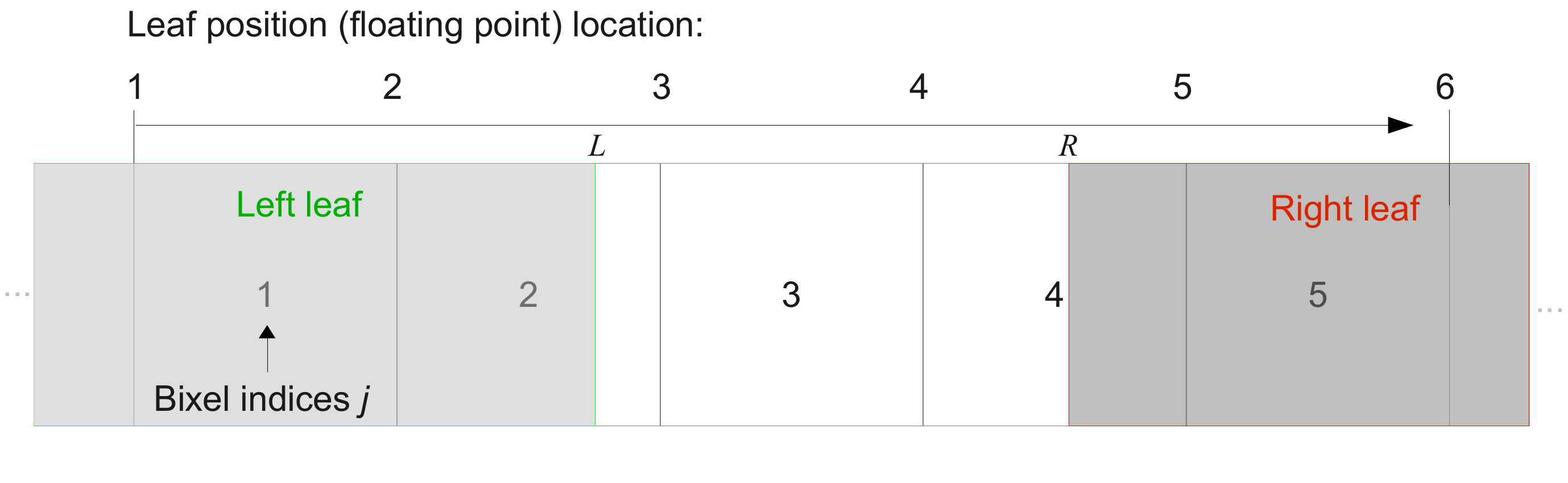}
\caption{Relationship between bixel indices and leaf positions. This illustration is for $B=5$ bixels. The leaves are drawn as gray rectangles and they are both shorter than they would actually be: the left leaf would extend leftward and the right leaf would extend rightward, as shown by the dots, for the opening shown. We assume that all leaves are physically long enough to cover the entire row of bixels if needed.}
\label{fig:bixels}
\end{figure}

Let $f_{ij}$ be the fluence map we are trying to produce. The index $i$ indexes the leaf rows and
$j$ indexes the columns (position along each row). We assume the bixels across a row are indexed 1, 2, $\ldots B$. Physical floating point leaf positions aligned with these bixel locations are such that ``1'' (e.g. centimeter) on the physical ruler corresponds with the left side of bixel with index $j=1$. Therefore, physical leaf positions for the left and right leaves range from 1 to $B+1$, see Figure \ref{fig:bixels}.

The key function used to map leaf positions to the fluence map produced is the exposure function $e$, which
gives the exposure (a number between 0 and 1) of beamlet $j$ given the left leaf position is $L$ and the right leaf position is $R$. This function is given as follows. The same function is applicable to all the rows, so there is no row index in the following:

\begin{align}
e(L,R,j) &= 
  \begin{cases}
    1  & \text{if } L\leq j \text{ and } R \geq j+1 \text{  [fully exposed]},\\
    j+1-L  & \text{if } j< L < j+1 \text{ and } R \geq j+1 \text{  [fully exp. by rt leaf, partially blocked by left]},\\
    R-j & \text{if }  L \leq j \text{ and } j < R < j+1 \text{  [fully exp. by left leaf, partially blocked by rt]},\\
    R-L & \text{if }  j < L < j+1 \text{ and } j < R < j+1 \text{  [partially blocked by both]},\\
    0 & \text{if }   L \geq j+1 \text{ or }  R \leq j \text{  [fully blocked by one of the leaves]}
  \end{cases}
\end{align}

We choose a fixed discretization for time, $\Delta$. For example $\Delta = 1/3$ of a second. Let $L^i_t$ be
the position of the $i$th leaf end at time step $t$, and $R^i_t$ is likewise defined. Note that if $L^i_t = R^i_t$, the leaves are closed. Let $D_t$ be the dose rate
at time $t$ (dose rate units are MU/sec, where MU stands for monitor units). The fluence map $g_{ij}$ (in MU) obtained by a set of leaf trajectories and dose rates is then given by:

\begin{equation}
g_{ij} = \sum_{t=1}^T e(L^i_t,R^i_t,j)D_t \Delta
\label{gij}
\end{equation}
\noindent where $T$ is the total number of time steps of the fluence map delivery.

The optimization problem of matching the fluence map $f$ is given by:
\begin{align}
\label{model1}
\text{min } & \sum_{i}\sum_{j} (f_{ij} - g_{ij})^2  \nonumber \\
\text{subject to:~~~~~~~~} g_{ij} &= \sum_{t=1}^T e(L^i_t,R^i_t,j)D_t \Delta \nonumber \\
L_t^i &\leq R_t^i,~~~~ \forall t,~i \nonumber \\
L_t^i - c \leq L_{t+1}^i & \leq L_t^i + c,~~~~ \forall t=1\ldots T-1,~i \nonumber \\
R_t^i - c \leq R_{t+1}^i & \leq R_t^i + c,~~~~ \forall t=1\ldots T-1,~i \nonumber \\
L_t^i & \geq 1,~~~~ \forall t,~i \nonumber \\
R_t^i & \leq B+1,~~~~ \forall t,~i \nonumber \\
0 \le D_t & \leq D_{max},~~~~ \forall t
\end{align}
\noindent where $c$ is a constant reflecting the maximum leaf speed constraint, and $e$ is the function given in (1). The $L,~R$ constraints are, in the order they appear: the left leaves must stay to the left of the right leaves, the left leaves cannot travel more than $c$ cm per time step (a typical maximum leaf speed is 3 cm/sec, so, with a time step of 1/3 sec, $c=1$), same for the right leaves, the left leaves should never go to the left of position 1, the right leaves should never go to the right of position $B+1$, and the dose rate at each time step should be non-negative and no larger than the maximum dose rate $D_{max}$. Note that we allow leaves to move back and forth rather than enforcing unidirectional 
motion across the field. We present a proof that unidirectional motion can be suboptimal in Appendix A.

\section{Solution approach}
Formulation \ref{model1} is a non-convex optimization model due to the non-convex mapping between leaf positions 
and beamlet exposure and the multiplication of the dose rate and leaf position variables. The non-convexity implies the existence
of local minima and therefore a global optimization procedure is needed. There is no finite time algorithm that solves general
non-convex optimization problems to proveable optimality. 
Gradient descent methods find local optimal solutions. We 
run gradient-based minimizations at a diverse set of starting solutions and choose the best overall solution. 
For the local minimizations, we use the Matlab (The MathWorks, Natick MA, version 7.14) function fmincon with the default
interior-point method with a user supplied gradient. The gradient computation is given in Appendix B. 

\subsection{Generating diverse starting solutions}
For generating diverse
starting feasible solutions, we randomize on the following trajectory types:

\begin{itemize} 
\item Type 1: Left to right leaf sweep
\item Type 2: Right to left leaf sweep
\item Type 3: Close-in trajectory
\item Type 4: Open-out trajectory
\item Type 5: Random leaf motion trajectory
\item Type 6: For each row, choose one of the above independently of the other rows.
\item Type 7: Left to right for large delivery time rows, close-in otherwise
\end{itemize}

Each of the above trajectories is randomly generated. For example, to generate a random feasible left to right
trajectory, we flip a coin at each time step to determine if a leaf will advance to the right at maximum speed or stay still.
One could use other distributions to randomly generate leaf steps as well. We ensure feasibility by making sure leaves do not
collide or extend out of range. Dose rates are set at $D_{max}$.

\subsection{Speeding up the optimization by focusing on tough rows}
\begin{figure}[!htbp]
\centering
\includegraphics[trim=70 170 70 150,clip,width=0.8\textwidth]{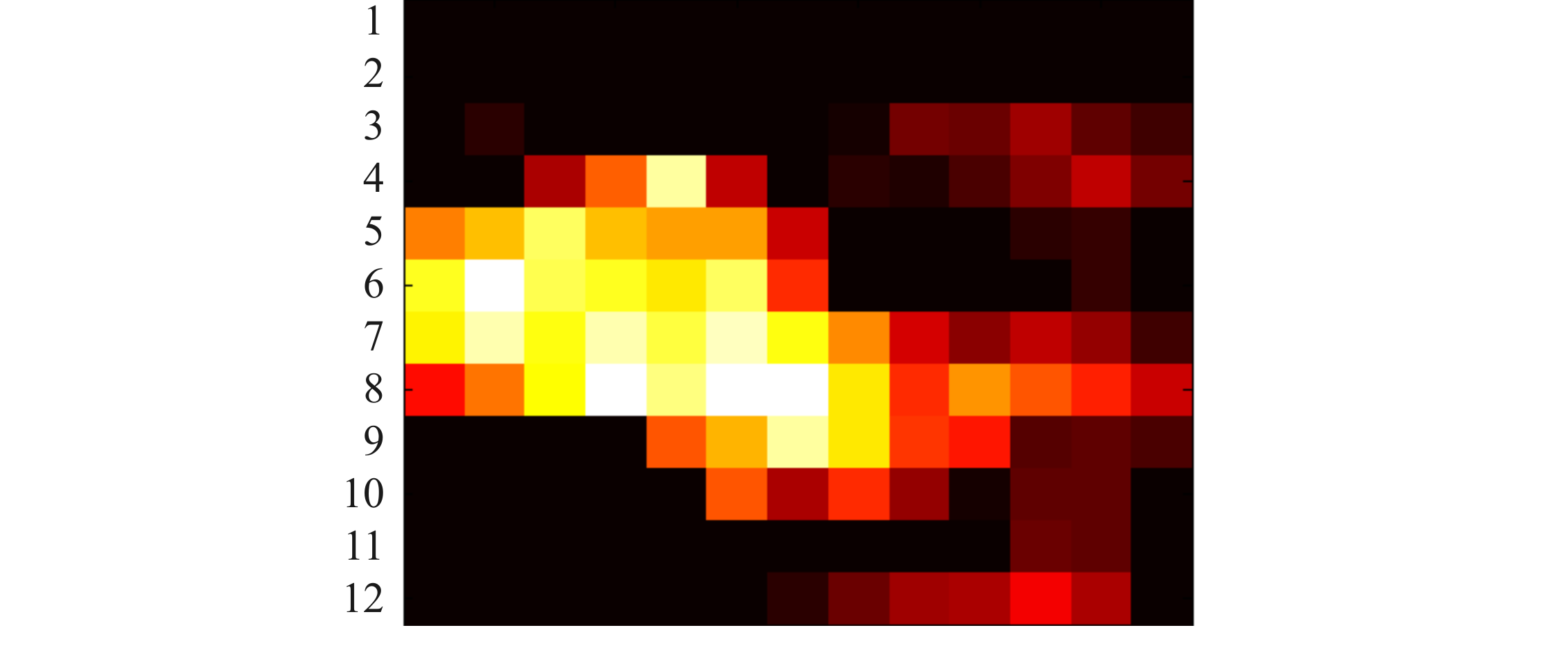}
\caption{An example of a fluence map where optimization is more difficult for some rows than for others.}
\label{fig:flMap6}
\end{figure}

The fluence often varies strongly among the rows of a fluence map, and for some rows it can be much more difficult to obtain the desired fluence than for others. This difference can be so large that for some rows a leaf trajectory that gives exactly the desired fluence can easily be found for any sequence of dose rates that is suitable for the tougher rows. Consider the fluence map in Figure \ref{fig:flMap6} as an example. Rows 7 and 8 have positive fluence over the full width of the row as well as a high total fluence, which means that they require a relatively large amount of delivery time in order to replicate the map properly. On the other hand, row 11 has a positive and low fluence in only two bixels. The amount of MUs required to deliver the fluence of rows 7 and 8 is thus much higher than what is required for row 11 and any sequence of dose rates that is suitable for rows 7 and 8 will suffice for the fluence in row 11 to be perfectly replicated.

The above observation allows us to decompose the solution approach into two steps. First, the dose rates and leaf positions are optimized for a reduced fluence map, from which the rows with a fluence that can easily be replicated are removed. This greatly reduces the number of variables, which is especially useful for non-convex optimization problems, and thus improves the solution times. In the second step, the dose rates are fixed to the values found in the first step, and the leaf trajectories for each of the easy rows are optimized individually. These are small and simple optimization problems for which only a small number of random starting points are required, so little time needs to be spent on finding a good solution. Decoupling the procedure in this manner cuts down the solution time by a factor of about three for maps of the size shown in Figure \ref{fig:flMap6} and an even larger factor for larger maps.

All rows with a very low total fluence compared to the other rows are considered easy. We define ``very low'' as any total fluence below 10\% of the maximum total fluence over all rows. Of the remaining rows, those with a high sum of positive gradients (SPG, \cite{craft-spg}) are considered to be tough, and the others easy. The SPG is the sum of all fluence increments over a row. If the SPG of a row is larger than the mean SPG over all rows, it is considered a difficult row. SPG is a good indicator of row delivery complexity since 
delivery time for a row using the sliding window technique increases linearly with SPG \cite{vmerge}.


\section{Results}

We demonstrate the method on two fluence maps, both generated from the data publically available via the CORT data set \cite{cort}. 
The first fluence map is from the prostate patient with lymph nodes and the second is from the head and neck patient.
For each fluence map we solve the complete fluence matching problem (i.e. generate many starting solutions and minimize each one using fmincon, then pick the overall best objective value) for several values of $T$ in order to generate the trade-off curve of delivery time and fluence map matching quality. A good upper bound on the maximum time needed for perfect fluence map delivery
is the maximum row delivery time for the leaf-sweep algorithm. The row delivery time is given as [time to sweep the leaves across the row at maximum leaf speed] + [SPG time], where [SPG time] is the SPG of the row (in MU) divided by the maximum dose rate (to convert the units into time) \cite{vmerge}.
For the prostate case this value is 6.3 seconds, and for the head and neck case it is 10.3 seconds.

\begin{figure}[!htbp]
\centering
\includegraphics[trim=0 0 0 0,clip,width=0.9\textwidth]{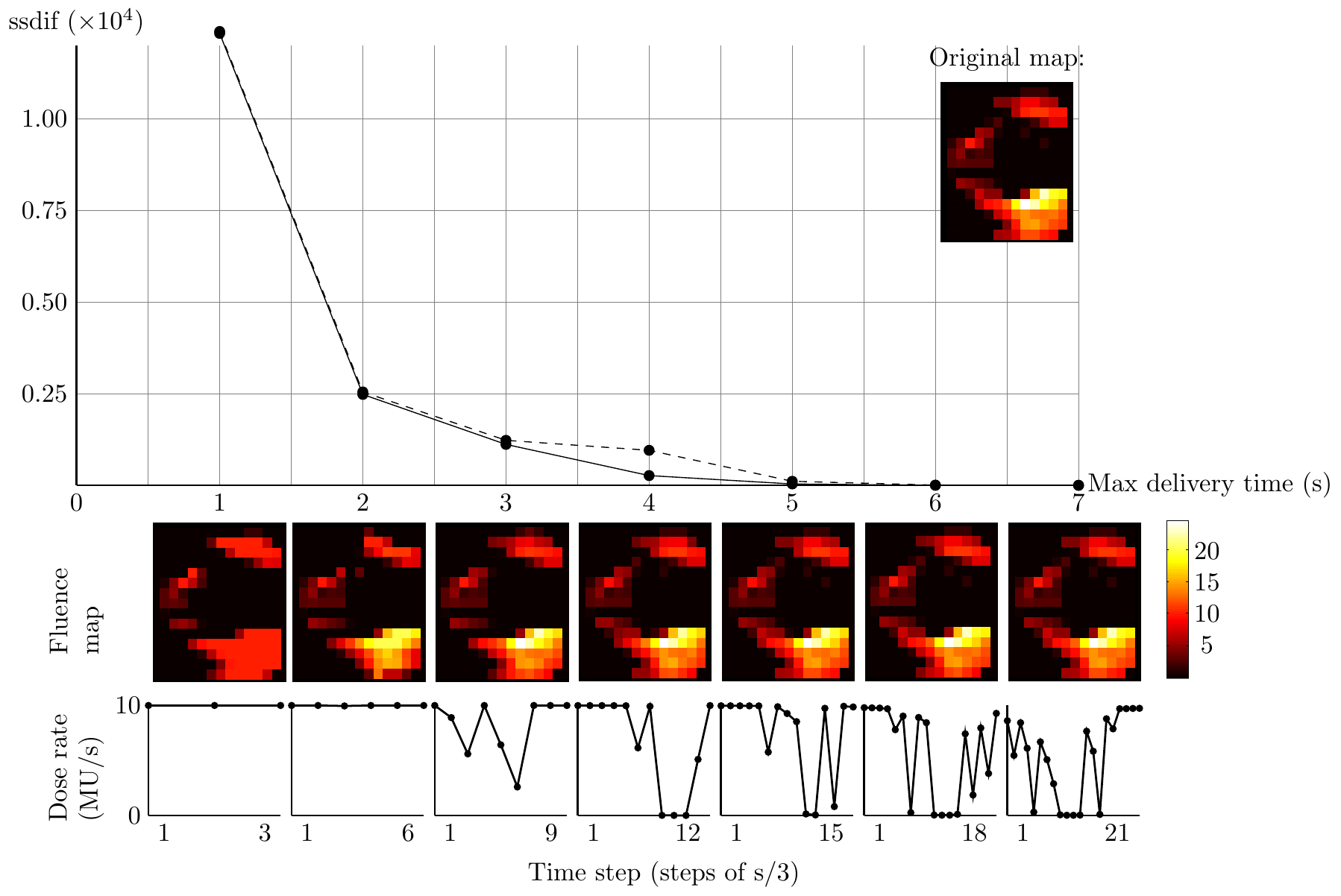}
\caption{Prostate case: Solution to the fluence map matching problem for various times. The y-axis of the graph shows the objective 
function that is minimized, $\sum_{ij} (f_{ij} - g_{ij})^2$, which is labelled as ssdif (sum of squared differences). Fluence 
map units are MU. The dose rate plots show the dose rate versus time for each delivery.
The solid line is for optimized dose rates and the dashed line is for fixed dose rate of 10 MU/sec (the maximum allowed value for
the optimized case).
Note that for very short times the maximum dose rate is used always due to the need to get in sufficient fluence.}
\label{fig:res1}
\end{figure}

As is typical in IMRT smoothing studies \cite{AlberNuesslin2000,craft-spg,diffsmooth}, 
we see that the fluence map can be delivered much faster with 
only neglible (indeed, barely visible) degradation, see Figure \ref{fig:res1}. Dose rates are seen to fluctuate in all cases except for the smallest total
allotted time case, where the dose rate remains at its maximum level in order to get enough fluence through. In all the other
cases, dose rates drop
in order to move leaves from one place to another without depositing dose. It is interesting to note that if given sufficient time, 
a fluence map can be perfectly delivered with a sliding window technique at constant maximum dose rate, but 
one can also achieve a near perfect match with a fluctuating dose rate. The dotted line shows the result for fixed dose rate of 10 MU/sec. For short time solutions, where using the maximum dose rate is optimal, there is no difference in the solutions, and for the large time solutions, where sliding window with maximum dose rate is also known to be an exact solutions, there is also no difference.

\begin{figure}[!htbp]
\centering
\includegraphics[trim=0 0 0 0,clip,width=0.9\textwidth]{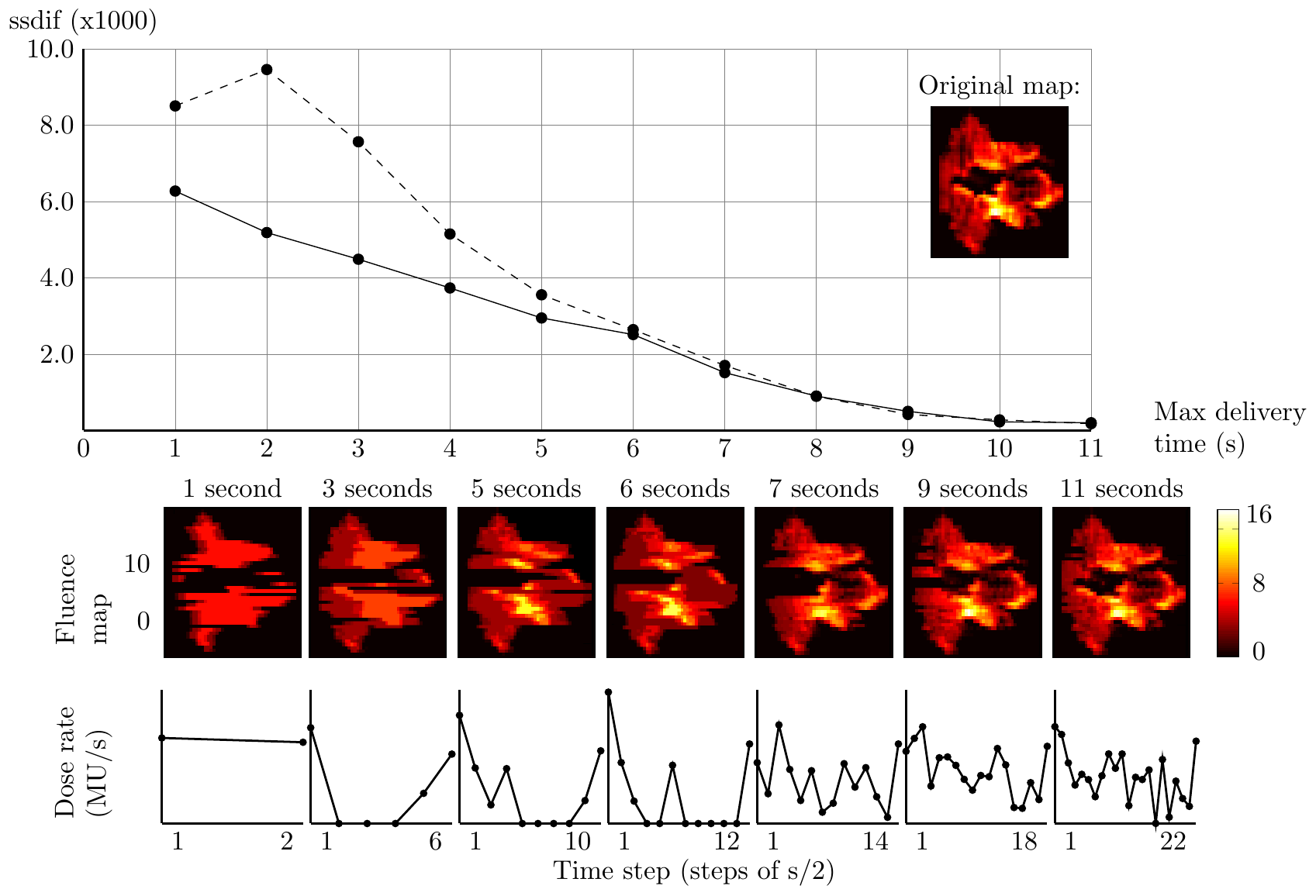}
\caption{Head and neck case, fluence map 17 (for description, see Figure \ref{fig:res1} legend).}
\label{fig:res2}
\end{figure}

\begin{figure}[!htbp]
\centering
\includegraphics[trim=0 0 0 0,clip,width=0.9\textwidth]{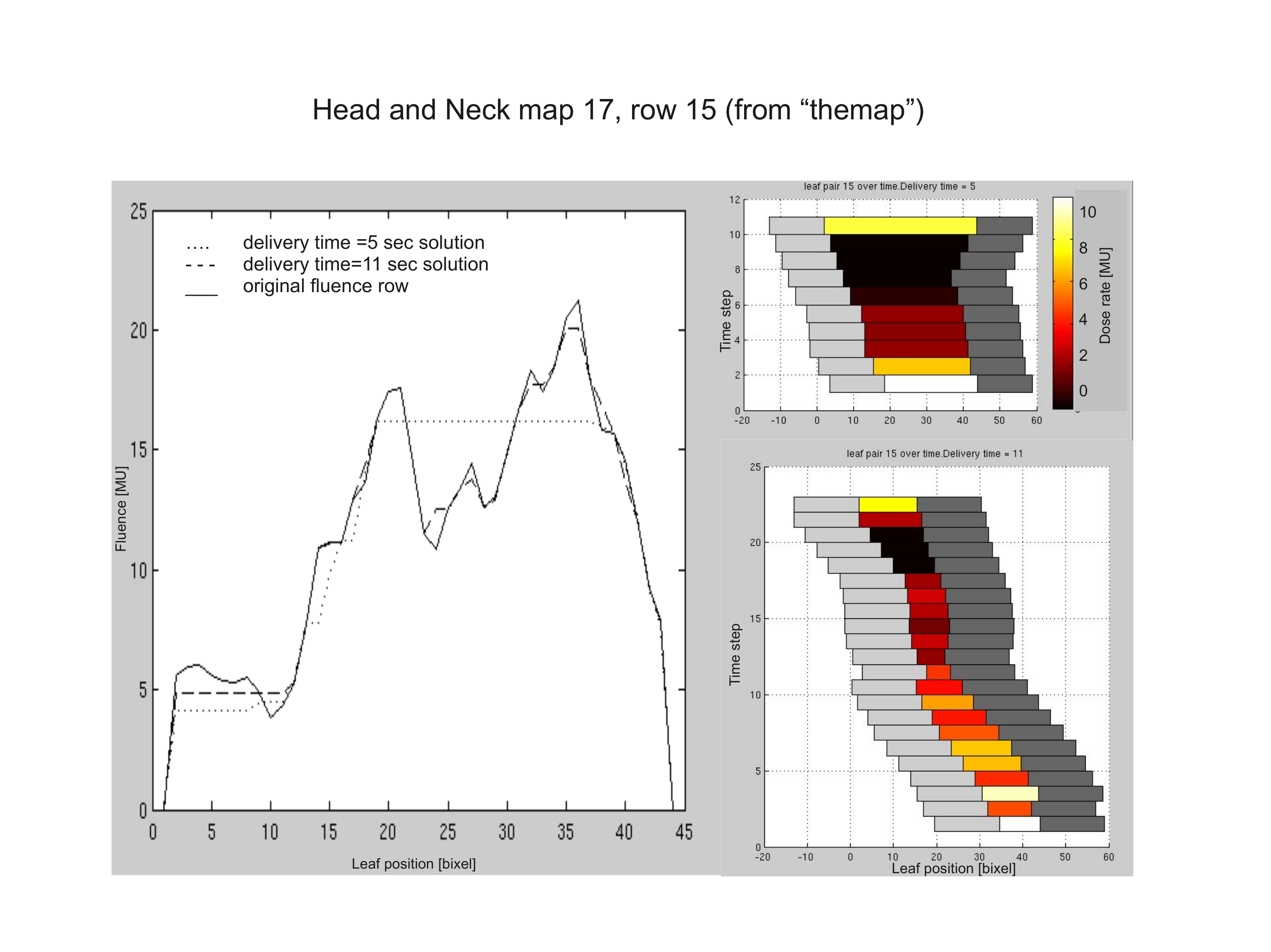}
\caption{Two sample leaf trajectories for leaf row 15 of head and neck map 17, which is a difficult (high SPG) row. The left plot shows
the fluence row that the optimizer is trying to match along with the resulting fluence from the optimal 5 second solution and the optimal 11 second solution. The right figures show these solutions and also indicate with the heat map the dose rate chosen at each time step (compare with dose rates as plotted
in Figure \ref{fig:res2}). In these heat maps, the lightest color, white, corresponds to the maximum dose rate of 10 MU/sec. The left leaf is depicted in light gray, the right leaf in dark gray.}
\label{fig:leaftraj}
\end{figure}

The head and neck case fluence maps are larger since the target size is larger and the beamlet size is 0.5cm $\times$ 0.5cm as compared with 1 cm$^2$ for the prostate case. A typical head and neck map from this data set has dimensions 40 leaf pairs $\times$ 44 columns. Figure \ref{fig:res2} depicts the trade-off curve between delivery time and fluence map matching quality for a head and neck case fluence map, with the original map being recreated shown in the top right corner. The optimized solutions, like for the prostate case, all show variable dose rates expect for the smallest time solution which uses the maximum dose rate the entire time. Figure \ref{fig:leaftraj} shows two optimal leaf trajectories for different total time allotted solutions from the trade-off curve. The fluence row that is chosen is a difficult row with three prominent peaks and smaller peaks between those. The 5 second solution, with the resulting fluence given by the dotted line, is clearly not enough time to replicate these peaks. The 11 second solution does very well, only slightly truncating the largest peak and approximating the small leftmost fluence peak with a flatter version. For the 11 second solution, the left leaf moves leftward, rightward, and then leftward again (i.e. the leaf trajectory is not unidirectional), and the right leaf uses a similar pattern. Both the 5 second and the 11 second solutions zero the dose rate towards the end in order to position the leaves for a final surge of radiation.

We investigate the sensitivity of the optimal solution for the head and neck case to the leaf speed.
Doubling the leaf speed from the default value of 3 cm/sec to 6 cm/sec produces a
dramatic improvement in the solution, with the objective function (the sum of squared differences) dropping from 2300 to 500. Further increasing the leaf speed to 9 cm/sec offers only marginal improvement to 460. Increasing the dose rate above the nominal value does not improve the solution at all in this regime, as can be gleaned from noting in the bottom of Figure \ref{fig:res2} that for total time = 7 seconds, the default dose rate is not an active constraint (the dose rate for this solution, while allowed to be as high as 10 MU/sec, rarely goes above 5 MU/sec). If the fluence map was globally scaled up (for example in a hypofractionated case where much more dose is delivered per treatment), then maximum dose rate would become an influential parameter.

\section{Discussion and conclusions}

We chose to formulate the dynamic leaf sequencing problem as a continuous optimization problem since simpler approaches,
such as manually determining leaf trajectories, proved too complex given a variable dose rate which couples all the leaf rows
together. Since the resulting formulation is necessarily non-convex, due in part to the fundamental step-function like 
relationship of leaf position and fluence transmission to a bixel, the solution procedure needs to have a global search aspect.
Global search algorithms such as particle swarm, simulated annealing, differential evolution, and  
genetic algorithms could be attempted, but
in order to keep the focus on the problem being solved, and knowing that gradient descent is a useful strategy for smooth
optimization problems, especially when an analytical gradient is available, we opted for a straightforward 
approach that could use this information. 
In order to gain confidence in the near-optimality of our approach, we let the random search and gradient descent optimizer
run for days for certain problem instances to show that after a few hours (problem size dependent of course) the solution
quality plateaus. Testing alternate global search strategies will be an interesting future study, although we note that
the search procedure used herein is easily parallelized, and the problem data size is small, so practical implementation
is already feasible.

The procedure described in this paper is applicable to static beam IMRT performed with dynamic delivery. For treatment 
planning systems such as Monaco, RayStation, and Pinnacle that perform IMRT optimization as a two step process: 1) fluence map computation and 2) leaf sequencing and refinement, the procedure can be used as the leaf sequencing procedure either for a single fixed time 
per fluence map or to generate a set of delivery time/plan quality trade-off curves as shown in Figures \ref{fig:res1} and \ref{fig:res2}.
Fluence map smoothing is also often incorporated into the fluence map optimization step, and how best to utilize both of these
smoothing techniques (i.e. directly incorporating smoothing into the optimization and implicitly smoothing in the sequencing procedure
decribed herein by choosing a small enough delivery time) is a topic for future investigation. 

Restricting large sudden changes in the dose rates can be done with linear constraints on the dose rate variable. 
Incorporating leaf gaps into the optimization can be done by altering the inequality $L_t^i \leq R_t^i$ with a given minimum clearance 
leaf gap, say 1 millimeter. If we assume that the jaws of the linac are fixed, leaf transmission can be accounted for by 
replacing the 0 term in the $e(L,R,j)$ equation, representing leaf blocking, with a transmission term. Moving jaws, 
either independent from the moving leaves or serving as a carriage on which the leaves move relative to, can also be modeled,
although the exposure function $e$ becomes more challenging. However, given the general modeling approach taken here, provided that for 
a given set of jaw and leaf positions an exposure function can be written down, we believe most or all linacs can be 
modeled with this technique.

Given that VMAT will likely become the dominant IMRT modality in the future, since it is more flexible than static beam IMRT and does not
``waste time'' moving the gantry without delivering dose, we believe that the main impact of this present work will be in its application
to VMAT planning. For this however, a significant challenge arises. In VMAT, a set of fluence maps optimized around the patient at a given 
angular frequency, say every 10 degrees, needs to be delivered. If each fluence map were sequenced individually as described here, 
than the leaf end
positions of one map would not be equal to (or even nearby) the leaf starting positions of the next map. Stopping the gantry
and turning off the dose rate to reposition the leaves for the next map would add much time to the total delivery and therefore
defeat the entire purpose of this work and the efficiency of VMAT. 
For the multi-map VMAT problem then, it is imperative to solve the fluence map sequencing
problems in a sequential coupled way. This is the subject of the second paper in this two-paper work.


\bigskip

\bibliographystyle{plain}
\bibliography{all}

\appendix

\section{The need for non-unidirectional leaf trajectories}

\begin{figure}[!htbp]
	\centering
	\subfloat[]{\includegraphics{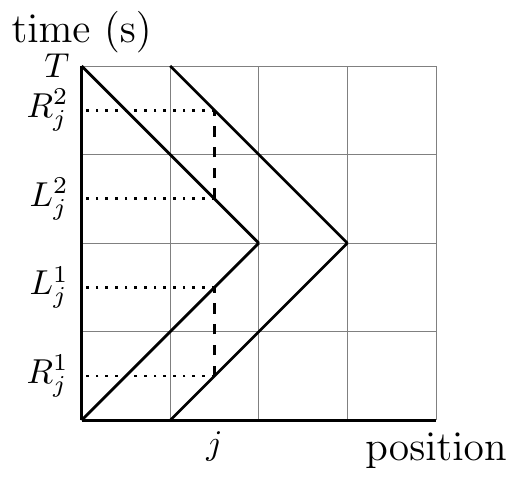}\label{fig:nonunidirectional2}}
	\quad\quad
	\subfloat[]{\includegraphics{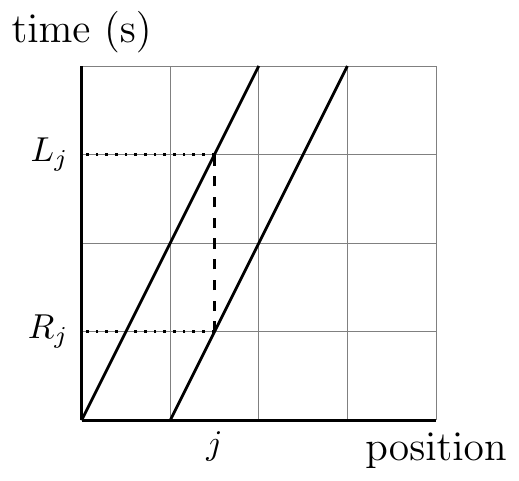}\label{fig:unidirectional}}
	\caption{An example of bidirectional (a) trajectories for the left and right leaf that can be transformed into unidirectional (b) trajectories without changing the fluence map.}
\end{figure}

Several works on treatment plan optimization for VMAT only consider unidirectional leaf trajectories, i.e., leaves can only move from left to right or right to left \cite{coupled,pappvmat}. If one uses a fixed dose rate, this is a suitable assumption, since any non-unidirectional leaf trajectory can be rewritten into a unidirectional trajectory without changing the fluence. In order to see this, consider the leaf trajectories in Figure \ref{fig:nonunidirectional2}, where the dose rate is constant. The total fluence deposited at position $j$ is indicated by the dashed lines, and is thus given by $L_j^1-R_j^1+R_j^2-L_j^2$, where $L_j^k$ and $R_j^k$ denote the time at which the left and right leaf, respectively, pass position $j$ for the $k^{\text{th}}$ time. The same fluence can be achieved with a unidirectional leaf trajectory, where $L_j=L_j^1-L_j^2$ and $R_j=R_j^1-R_j^2$. This gives the unidirectional leaf trajectory in Figure \ref{fig:unidirectional}. The fluence deposited with the unidirectional trajectories is $L_j-R_j = L_j^1-L_j^2 - R_j^1+R_j^2$, which is equal to the fluence deposited in the bidirectional case. This approach can be extended to nonunidirectional maps where the leaves move in more than two, say $n$, directions by using the transformation $L_j=L_j^1-L_j^2+...+L_j^n$ if $n$ is even, and $L_j=L_j^1-L_j^2+...+T-L_j^n$ if $n$ is odd, and $R_j$ is likewise defined. Note that the above transformations only hold when the leaves first move from left to right, which can be assumed without loss of generality.

The constraints on leaf positions and leaf speed remain valid. For leaf speed, this is intuitive: the time a leaf uses to move from one position to another will never be reduced, but it may be increased. Thus, if the multidirectional plan satisfies the leaf speed constraints, then the unidirectional plan satisfies those as well. In order to see that the left leaf remains on the left of the right leaf, we need to show that the right leaf traverses a position $j$ before the left leaf does, i.e., $L_j>R_j$. Recall that we assume the leaves first move from left to right, then from right to left, and so on, which implies $R_j^1<L_j^1$, $R_j^2>L_j^2$, $R_j^3<L_j^3$, etc. This implies that $L_j=L_j^1-L_j^2+...$ is at least $R_j=R_j^1-R_j^2+...$, and hence $L_j>R_j$.

\begin{figure}
	\begin{minipage}[b]{0.45\textwidth}
		\centering
		\includegraphics[width=0.65\linewidth]{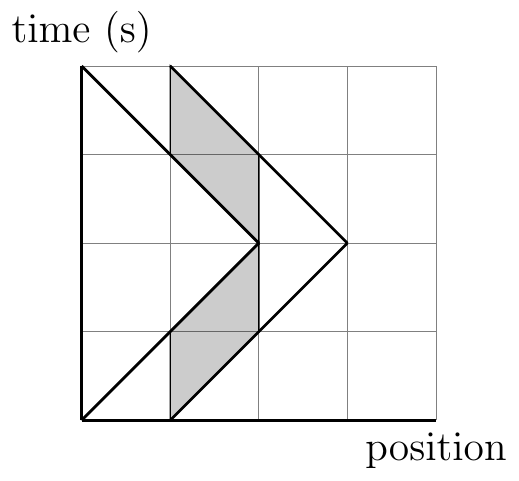}
		\caption{The total fluence delivered to a bixel is equal to the area enclosed by the leaf trajectories. \label{fig:surfaceFluence}}
	\end{minipage}
	\quad\quad
	\begin{minipage}[b]{0.5\textwidth}
		\centering
		\includegraphics[width=\linewidth]{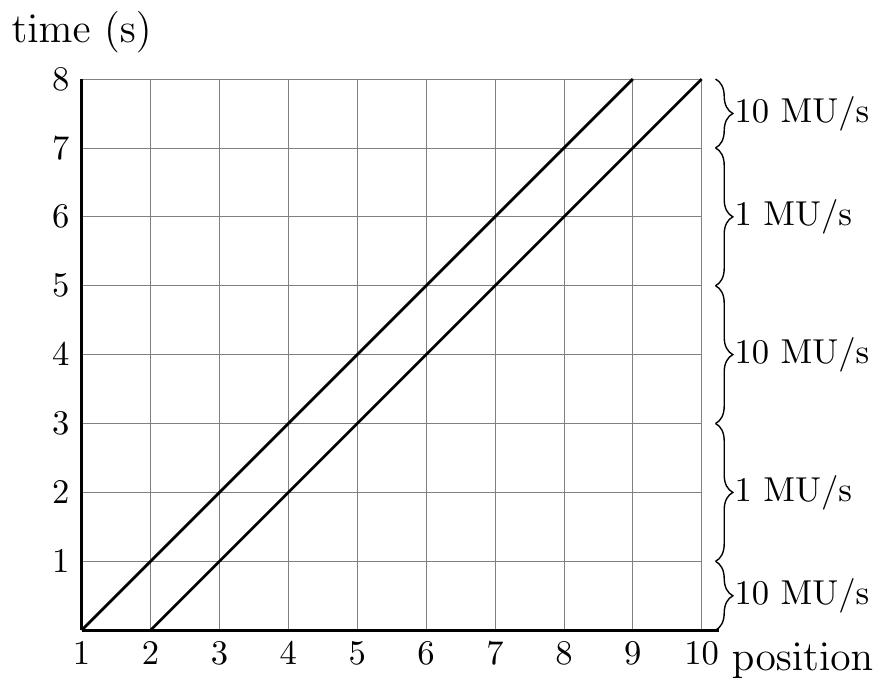}
		\caption{The only optimal VMAT treatment plan for a leaf pair whith desired fluence map [5 5.5 1 5.5 10 5.5 1 5.5 5] when time is limited to $T=8$.}\label{fig:leftrightmap}
	\end{minipage}
\end{figure}

When the dose rate is allowed to vary, unidirectional leaf trajectories may not be optimal as is shown in the following example. Keep in mind that the fluence at a bixel is visible in a leaf trajectory graph as the surface enclosed by the leaf trajectories and the boundaries of the bixel, see Figure \ref{fig:surfaceFluence} where the fluence to the second bixel is equal to the surface of the grey area. Suppose that for a given maximum number of time steps $T=8$, we aim to find leaf trajectories that yield the fluence map [5 5.5 1 5.5 10 5.5 1 5.5 5] for the first leaf pair. The low fluence in bixels 3 and 7 can either be achieved by letting the leaves pass these bixels at maximum leaf speed with a small distance between them, or by lowering the dose rate. Moving the leaves past these bixels in close proximity to each other is not possible due to the time restriction, so turning the dose rate down is the only possibility. This gives the leaf trajectories and dose rates as shown in Figure \ref{fig:leftrightmap}.

\begin{figure}[!htbp]
	\centering
	\subfloat[]{\includegraphics[width=0.45\textwidth]{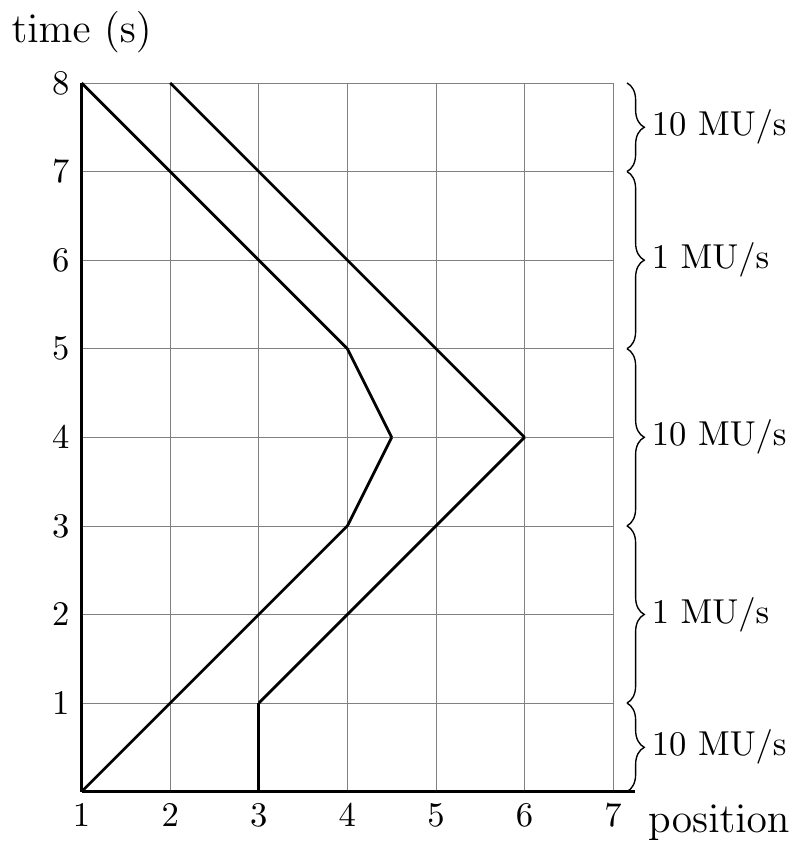}\label{fig:secondLeafPair2nonUni}}
	\quad\quad
	\subfloat[]{\includegraphics[width=0.45\textwidth]{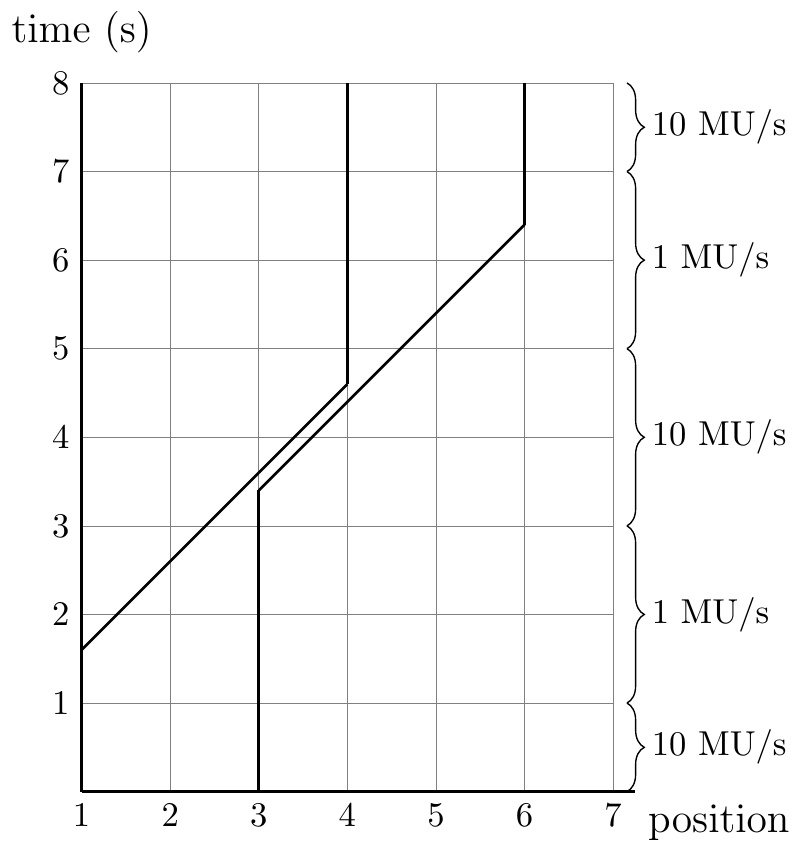}\label{fig:secondLeafPair2Uni}}
	\caption{Figure (a) shows a non-unidirectional VMAT plan that gives fluence map [10 17 2 17 10 0 0 0 0]. With the given dose rates, it is not possible to find unidirectional leaf trajectories that yield the fluence map [10 17 2 17 10 0 0 0 0], as is shown in (b). }
\end{figure}

Now suppose we have another leaf pair with fluence map [10 17 2 17 10 0 0 0 0]. Given the dose rates enforced by the first leaf pair, there exist non-unidirectional leaf trajectories that can perfectly replicate this map (see Figure \ref{fig:secondLeafPair2nonUni} for an example). This fluence cannot be achieved with unidirectional leaf trajectories for the given dose rates, which we show by aiming to obtain such a leaf trajectory. Due to the symmetry, we deliver the dose to the third bixel halfway the time period by letting the leaves pass bixel 3 at maximum leaf speed and with a distance of 0.2, allowing for a delivery of 2 MU. Note that the fluence to this bixel cannot be delivered in one of the time periods with low dose rate, since this does not allow enough time to deliver the fluence in bixels 1 and 2 or 4 and 5. Given the leaf trajectory for bixel 3, we let the left leaf pass bixels 1 and 2 as late and as fast as possible, and let the right leaf start at the end of bixel 2 (see Figure \ref{fig:secondLeafPair2Uni}). This gives the maximum possible dose to bixel 2, which is lower than the desired fluence. Thus, a perfect replication of the fluence map cannot be delivered with unidirectional leaf trajectories.

\section{Gradient and Hessian}

The gradient of the objective function $F = \sum_{ij} (f_{ij} - g_{ij})^2$, 
where $g_{ij}(L_t^i,R_t^i,D_t)$ is substituted by its definition, is given by:
\begin{align}
\frac{\partial F}{\partial D_t} & = -2\sum_{i,j} (f_{ij} - \sum_{s=1}^T e(L_s^i,R_s^i,j) \Delta D_s)e(L_t^i,R_t^i,j)\Delta \\
\frac{\partial F}{\partial L_t^i} & = -2\sum_{i,j} (f_{ij} - \sum_{s=1}^T e(L_s^i,R_s^i,j) \Delta D_s)e'_{L}(L_t^i,R_t^i,j)\Delta D_t\\
\frac{\partial F}{\partial R_t^i} & = -2\sum_{i,j} (f_{ij} - \sum_{s=1}^T e(L_s^i,R_s^i,j) \Delta D_s)e'_R(L_t^i,R_t^i,j)\Delta D_t.
\end{align}
In the above, the functions $e'_{L}(L_t^i,R_t^i,j)$ and $e'_{R}(L_t^i,R_t^i,j)$ are the derivatives of $e(L_t^i,R_t^i,j)$ with respect to $L$ and $R$, respectively:
\begin{align}
e'_L(L,R,j) & = \left\{ \begin{array}{ll} 0 & \text{ if } L<j \text{ and } R>j \text{, or } R \leq j \text{ or } L>j+1 \\[0pt]
-1 & \text{ if } j<L<j+1 \text{ and } R>j \\[0pt]
[-1,0] & \text{ if } L=j \text{ or } L=j+1 \end{array} \right.
\end{align}
\begin{align}
e'_R(L,R,j) & = \left\{ \begin{array}{ll} 0 & \text{ if } L<j+1 \text{ and } R>j+1 \text{, or } L \geq j+1 \text{ or } R<j \\[0pt]
1 & \text{ if } L<j+1 \text{ and } j<R<j+1 \\[0pt]
[0,1] & \text{ if } R=j \text{ or } R=j+1 \end{array} \right.
\end{align}

From this, we can compute the Hessian using the following second order derivatives:
\begin{align}
\frac{\partial^2 F}{\partial D_t^2} & = 2 \Delta^2\sum_{i,j} e(L_t^i,R_t^i,j) \\
\frac{\partial^2 F}{\partial (L_t^i)^2} & = 2 \Delta^2 \sum_{i,j} e'_L(L_t^i,R_t^i,j)^2 D_t^2 \\
\frac{\partial^2 F}{\partial (R_t^i)^2} & = 2 \Delta^2 \sum_{i,j} e'_R(L_t^i,R_t^i,j)^2 D_t^2 \\
\frac{\partial^2 F}{\partial D_t \partial L_t^i} & = 2 \Delta \sum_{i,j} e'_L(L_t^i,R_t^i,j) \left( g_{ij} - f_{ij} + e(L_t^i,R_t^i,j) D_t \right) \\
\frac{\partial^2 F}{\partial D_t \partial R_t^i} & = 2 \Delta \sum_{i,j} e'_R(L_t^i,R_t^i,j) \left( g_{ij} - f_{ij} + e(L_t^i,R_t^i,j) D_t \right) \\
\frac{\partial^2 F}{\partial L_t^i \partial R_t^i} & = 2 \Delta^2 \sum_{i,j} e'_L(L_t^i,R_t^i,j) e'_R(L_t^i,R_t^i,j) D_t^2.
\end{align}

The solution time of the interior-point optimization can be strongly reduced by giving the gradient to the solver. However, in our case supplying the Hessian does not yield improvements. We observed that reducing the time the solver spends on computing the Hessian yields an overall time speed-up, indicating further that the Hessian is not helpful for this optimization problem. We include the Hessian calculation for reference and in the event that is useful for other optimization approaches.


\end{document}